\documentclass[11pt,prd,aps,nofootinbib,floats,floatfix,amsmath,amssymb,secnumarabic]{revtex4-1}
\usepackage{amsmath}
\usepackage{amsfonts}
\usepackage{amssymb}
\usepackage{feynmp}
\usepackage{latexsym}
\usepackage{graphicx}
\usepackage[english]{babel}
\usepackage{listings}
\usepackage{natbib}
\usepackage{url}
\newcommand{\be}{\begin{equation}}
\newcommand{\ee}{\end{equation}}
\newcommand{\ba}{\begin{array}}
\newcommand{\ea}{\end{array}}
\newcommand{\bqa}{\begin{eqnarray}}
\newcommand{\eqa}{\end{eqnarray}}
\renewcommand{\d}{\mathrm{d}}

\renewcommand{\Re}{\mathbb{R}\textnormal{e}}
\renewcommand{\Im}{\mathbb{I}\textnormal{m}}

\newcommand{\e}[1]{e^{#1}}

\newcommand{\IRdiv}{\text{IR divergence}}

\begin{document}

\title{IR Divergences in Inflation and Entropy Perturbations}

\author{Wei Xue$^1$\footnote{xuewei@hep.physics.mcgill.ca}, Xian Gao$^{2,3,4}$\footnote{xgao@apc.univ-paris7.fr} and
Robert Brandenberger$^1$\footnote{rhb@physics.mcgill.ca}}

\affiliation{1) Department of Physics, McGill University,
3600 Rue Universit\'e, Montr\'eal, Qu\'ebec, Canada H3A 2T8}

\affiliation{2) Astroparticule \& Cosmologie, Universit\'{e} Denis Diderot-Paris 7,
        10 rue Alice Domon et L\'{e}onie Duquet, 75205 Paris, France}
        
\affiliation{3) ${\mathcal{G}}{\mathbb{R}}
\varepsilon{\mathbb{C}}{\mathcal{O}}$, Institut d'Astrophysique de Paris, Universit\'{e} Pierre et Marie Curie-Paris 6, 98bis Boulevard Arago, 75014 Paris, France}

\affiliation{4) Laboratoire de Physique Th\'{e}orique, \'{E}cole Normale Sup\'{e}rieure, 24 rue Lhomond, 75231 Paris, France}

\begin{abstract}

We study leading order perturbative corrections to the two point
correlation function of the scalar field describing the curvature perturbation
in a slow-roll inflationary background, paying particular
attention to the contribution of entropy mode loops. We find that
the infrared divergences are worse than in pure de Sitter space:
they are power law rather than logarithmic. The validity of perturbation
theory and thus of the effective field theory of cosmological
perturbations leads to stringent constraints on the coupling constants
describing the interactions, in our model the quartic
self-interaction coupling constant of the entropy field. If the
self coupling constant is larger than some critical value which
depends in particular on the duration of the inflationary phase,
then perturbation theory breaks down. Our analysis may have
implications for the stability of de Sitter space: the
quantum effects which lead to an instability of de Sitter space will
be larger in magnitude in the presence of entropy fluctuations.

\end{abstract}

 \maketitle

\newpage

\section{Introduction}

The inflationary universe scenario provides an explanation for the  observed inhomogeneities
in the distribution of galaxies and  the anisotropies of the cosmic microwave background.
According to inflationary cosmology, the primordial fluctuations arise from initial
vacuum  perturbations of the canonical variable which describes the coupled metric
and matter inhomogeneities ( see~\cite{Mukhanov:1990me} for a overview of cosmological
perturbation theory). The standard computations are done at tree level.
However, since Einstein's gravitational theory is non-linear, it is not valid to neglect
interactions which appear at higher order in perturbation theory. Such interactions
will lead to non-Gaussianities  in the spectrum of cosmological perturbations, a topic
which has received a lot of recent interest (see e.g. ~\cite{Bartolo:2001cw} for an
early reference). Interactions also effect the two point functions via loop
corrections during inflation (see e.g. ~\cite{Weinberg:2005vy}, and they can even
lead to back-reaction effects on the cosmological background itself, see e.g.
\cite{Woodard, ABM} for early works computing the effects of graviton and scalar
loops, respectively).

In order to study quantum field theory during inflation or, more ambitiously, to understand
quantum gravity, it is necessary to systematically treat perturbation theory. As a non-renormaliziable
theory, General Relativity needs an infinite number of counter-terms to get rid of the
ultraviolet (UV) divergences (see e.g. the textbook of \cite{Birrell} and also the more recent
articles \cite{Weinberg:2005vy,Senatore:2009cf}).
Since in inflationary cosmology it is assumed that UV modes originate in their vacuum state,
the contribution of UV physics to loop corrections of correlation functions is tiny and beyond
the interest for observations. On the other hand, the contributions of infrared (IR)
modes to correlation functions can be large in inflationary cosmology due to the large
phase space of such IR modes. It is the Hubble radius $H^{-1}$ which separates UV
from IR scales. The Hubble radius is approximately constant (in physical coordinates)
during the phase of inflationary expansion. For a massless field there are both
UV and IR divergences. If we employ a brute cutoff renormalization scheme to deal with the two
kinds of divergences, then the UV cutoff scale must be taken to be fixed in physical coordinates,
while the IR cutoff is a comoving one~\cite{Xue:2011hm}. Hence, during a period of accelerated
expansion of space the phase space of IR modes increases. Thus, in contrast to
UV physics, IR physics may well be lead to observationally relevant effects
(see e.g. \cite{Woodard, ABM, Giddings:2010nc, Giddings:2011zd} since the long
distance modes can back-react on the semi-classical space-time or on other
gravitational perturbation modes. There have been a lot of studies on the
effects of infrared divergences on inflationary observables (see \cite{Seery:2010kh} for
a recent review and e.g. \cite{IRstudies} for some original studies).
There is an emerging consensus that in the presence of only adiabatic fluctuations,
the effect on cosmological observables will be small \cite{Giddings:2011zd, small}. However, as stressed
in \cite{Ghazal2} in the context of studies of the effect of back-reaction of the infrared
modes on the background cosmology itself, the story changes completely if entropy
modes are present. Since matter contains many fields, it is not realistic to neglect the
presence of these entropy modes.

There is a second aspect of IR problems in cosmology:
As pointed out in \cite{Linde:1980ts}, perturbation theory may break down because of
the contribution of IR modes to loop integrals. In thermal field theory this is called
the ``Linde problem": In the thermal ensemble the soft gluons contributing as IR loops to
Feynman diagrams cause the complete failure of perturbation theory since in fact the transverse
gluons obtain a mass via screening. In de Sitter space-time, a scalar field with a mass
which is tiny compared with the Hubble scale will cause the same problem as in thermal
field theory (see e.g. ~\cite{Burgess:2010dd} and also \cite{Riotto}). For an exactly  massless field, the
IR loops will be logarithmically divergent \footnote{See also \cite{Losic} for an
independent study of how infrared divergences in inflationary cosmology can lead
to a breakdown of cosmological perturbation theory.}.

The analysis of infrared divergences in inflationary cosmology is complicated by the
fact that we must treat both metric and matter fluctuations since they are coupled via
the Einstein equations. This mixing must be taken into account in the computation
of observables. The observable of particular interest is the curvature fluctuation
variable $\zeta$ \cite{Bardeen, Mukhanov:1990me}. In particular, in the case
of a massive matter field in the case of inflationary cosmology it turns out that the tilt
of the curvature perturbations and of the gravitational wave spectrum is similar to what
is obtained in the case of a massive scalar field in  pure de Sitter space-time - thus
there is a power law IR divergence in inflationary space-time.

In this paper, we analyse the IR divergences arising via loop effects in the two
point correlation function of the curvature fluctuation variable $\zeta$ in the case
of an inflationary space-time. We find a power law divergence rather than the
logarithmic divergence which appears in pure de Sitter space-time. The IR
divergences can be regulated, but the regulator is related to the initial condition at
the beginning of the phase of inflationary expansion. Field theory in an
expanding space-time is different from quantum field theory in Minkowski space-time:
As the inflationary phase begins, the vacuum keeps generating UV modes deep inside the
Hubble radius, modes which exit the Hubble radius, freeze out and are subsequently
squeezed and become the IR ensemble of quasi-classical fluctuations which will
back-react on the background \cite{Woodard, ABM} and impact the newly-generated
gravitational modes. Compared with the mild logarithmic divergence of pure de
Sitter space-time, the power law IR divergences in inflationary cosmology will put
stronger constraints on the applicability or perturbative methods in inflation.

However, it has been shown that the leading order contributions of IR divergences
in inflationary cosmology cancel, which can be proven by applying either the in-in
formalism~\cite{Seery:2010kh} or by semi-classical relations~\cite{Giddings:2010nc}.
Also, in the case of pure adiabatic fluctuations, IR divergences from naive curvature perturbations
can be locally gauged away (see e.g. \cite{Unruh, Abramo, Ghazal1, Yuko}), while
in the presence of entropy perturbations the IR effects are locally measurable and can no
longer be gauged away. Thus, the divergences from entropy modes are
physically relevant \cite{Ghazal2}. Meanwhile, the entropy perturbations are the source of  curvature
perturbations and, more specifically, lead to the growth of curvature perturbations on
super-Hubble scales. Thus, IR divergences will effect the observed curvature fluctuations.
Hence, the fact that the curvature fluctuations are time-dependent, the presence of
entropy perturbations, and the stronger IR divergences from loop effects will change
the story of IR divergences in inflationary cosmology compared to what happens in the
case of a spectator scalar field in de Sitter space.

In this paper, we will study the power law IR divergences and the entropy perturbations
in inflationary cosmology. In  Sec.~\ref{dotzeta=0}, we will discuss the power law divergences
arising in single field inflation models. This effect is general, and can be applied to multiple
field models as well. The new feature in multi-field models is the presence of entropy fluctuations
which we study in the subsequent section, We first consider  massless entropy modes,  and
then  massive ones. We derive  non-trivial bounds on the magnitude of the self-coupling
of the entropy modes which have to be satisfied in order for perturbation theory to remain
applicable. Finally, we study the case of scalar matter fields which are not coupled to the inflaton. 
However, there is a coupling with gravity, it this also results in constraints on the self coupling
constant if perturbation theory is to be reliable.

\section{Case of Conserved $\zeta$}
\label{dotzeta=0}

In this section, we compute the leading loop corrections to the two point function of the
curvature fluctuation $\zeta$ for single field inflation, in which case $\zeta$ is conserved
on super-Hubble scales. As we will show a bit later, the most convenient gauge for the
loop calculations is  comoving gauge. We take into account that the Hubble parameter
is not exactly constant during inflation. We show that this has an essential effect on
IR divergences. Our discussion is based on describing space-time using Einstein's
gravitational action, and considering matter to be composed of a set of minimally
coupled scalar fields.

\subsection{Single Field Inflation}

Single field inflation is obtained by using the condensation of a single scalar field
$\phi$ with a non-trivial but very flat potential in order to drive space to expand nearly
exponentially.  The resulting space-time background is often called
``(quasi)-de Sitter space", The background metric can be written in the form
\be
\d s^2= - \d t^2 + a^2(t) \d x^2 = a^2(\eta) \left( - \d \eta^2 + \d x^2 \right)  \ ,
 \ee
where  $t$ is physical time, $x$ are the comoving spatial coordinates, $\eta$ is the
conformal time, and the scale factor $a(t)$ takes the form
$a(t) = \e  {H t} = -\frac{1}{H\eta}$ in the case of pure de-Sitter space.

When introducing gravitational fluctuations, it is important to keep in mind that gravity
is similar to a gauge field theory in that there is a redundant local symmetry, namely
space-time diffeomorphism invariance. Out of the ten naive metric degrees of
freedom, there are only six physical ones - two scalar, two vector and two tensor
(see \cite{Mukhanov:1990me} for an in-depth discussion). The two tensor modes
are the two polarization states of the gravitons, in an expanding universe the two
vector modes can be neglected, and the two scalar metric and one matter degrees
of freedom are constrained by two of the Einstein equations, leaving one physical mode.
To get rid of the redundancy in the scalar sector, we can choose different gauges,
but physics will not depend on our gauge choices. Uniform spatial curvature gauge
is obtained by choosing the time slices such that the scalar metric perturbation (the
component of the metric corresponding to the scalar mode) vanishes. In this
case the scalar matter field perturbation will be non-vanishing. The spatial
part of the metric takes the form
\be
  g_{ij} = a^2\left(\delta_{ij} +h_{ij}\right) \ ,
\ee
where $h_{ij}$ is a transverse and traceless tensor representing the gravitons.

The other convenient gauge in the case of single field inflation is the comoving gauge
where the matter field fluctuation $\delta \phi$ is set to zero and the perturbation
appears as a fluctuation of the spatial curvature:
\bqa \label{metric}
 \delta \phi \, &=& \, 0  \nonumber \\
g_{ij} \, &=& \,  a^2 \left[ (1+2\zeta) \delta_{ij} + h_{ij} \right] \ .
\eqa
The remaining degrees of freedom of the perturbations in the specific time slice are
curvature perturbation $\zeta$ and graviton $h_{ij}$. As we will argue below, for our
purposes the comoving gauge has advantages.

At quadratic order, scalar metric fluctuations and gravitons decouple.
In the comoving gauge, the quadratic actions of curvature perturbations and gravitons
are as follows
\be
S_{2,s} = \frac{1}{2} \int \d t \d x^3 \epsilon \left[ a^3 \dot \zeta^2 - a (\partial \zeta)^2 \right]
\ee
and
\be
S_{2,t} = \frac{1}{8} \int \d t\d x^3  \left[ a^3 \dot h_{ij}^2 - a \left(\partial_l  h_{ij} \right)^2 \right] \, ,
\ee
where the subscripts $s$ and $t$ stand for the scalar and tensor contributions, respectively.
In the above, we are working in Planck units, i.e. to obtain the dimensionless metric
components of (\ref{metric}), we must divide the values of $\zeta$ and $h$ appearing in
the above action by the Planck mass $m_{pl}$.

At quadratic order, each Fourier mode evolves independently.
The solutions of the equations of motion for $k$ modes normalized to
correspond to quantum vacuum perturbations on sub-Hubble scales are
\be
 \zeta_k = \frac{H}{ \sqrt{2\epsilon}  \sqrt{2 k^3} } ( 1+ i k \eta) e^{-i k\eta}
\ee
and
\be
h_{\pm , k} = \frac{H}{  \sqrt{ k^3} } ( 1+ i k \eta) e^{-i k\eta} \, .
\ee
Here $\epsilon$ is the first slow-roll parameter, which is defined as the rate of
change of the Hubble scale in one Hubble time:
\be
\label{sr1}
\epsilon = -\frac{\dot H} {H^2} = \frac{\dot \phi^2}{2H^2} \ .
\ee

In de Sitter space, the Hubble scale is constant, while during inflation, the Hubble scale
is decreasing since the inflaton is slowly rolling down its potential. The first slow roll parameter
is required to be very small during the inflationary phase, and moreover, must be kept small
for the duration of inflation which corresponds to at least $60$ e-foldings. Hence there
are other parameters which characterize the rate of change of $\epsilon$, and  they are
also required to be small. The second slow-roll parameter is
\be
\eta \equiv \frac{V''}{V} \sim -\frac{\ddot \phi}{H \dot \phi} + \frac{1}{2} \frac{\dot \phi^2}{H^2}  = 2 \epsilon - \frac{\dot \epsilon}{H \epsilon} \ .
\ee
Sometimes we require inflation to last more than 60 e-foldings, such that $\epsilon$ is slowly
changing with $ \frac{\dot \epsilon}{H \epsilon}  \sim \mathcal{O} (\epsilon^2)$ which is of
next order of $\epsilon$. In this case $\eta \simeq 2 \epsilon$.

Although the Hubble scale is not exactly constant,
$\zeta$ and $h$ are good quantities to describe the perturbations in inflationary cosmology
since they are conserved outside the Hubble radius as long as there are no entropy
perturbations. The reason that they are conserved is the rescaling symmetry of the equations,
as can be seen from the action which is total derivative outside the horizon, such that the
curvature and graviton perturbations are not dynamical at all after Hubble radius crossing.

Therefore, the amplitudes of $\zeta_k$ and $h_k$ for modes whose wavelengths are larger
than the Hubble radius, can be described by the Hubble scale and the slow roll parameters at
the time when the modes cross the Hubble radius:
\be
 \zeta_k = \frac{H_*}{\sqrt{2\epsilon_*}  \sqrt{2 k^3} } ( 1+ i k \eta) e^{-i k\eta}
\ee
and
\be
h_{\pm , k} = \frac{H_*}{ \sqrt{ k^3} } ( 1+ i k \eta) e^{-i k\eta} \ ,
\ee
where subscripts $*$ mean that the corresponding quantities are evaluated at the Hubble
crossing time. Although on super-Hubble scales curvature perturbations and gravitons do
not depend on time, they are functions of $k$ because inflation is not exactly de Sitter.
\bqa
\langle \zeta_k \zeta_k \rangle \, &\simeq& \, \frac{H^4}{2k^3 \dot \phi^2}  \\
\langle h_k h_k \rangle \, &\simeq& \, \frac{H^2}{k^3 } \, .
\eqa

The $k$ dependence of the perturbations is characterized by spectral indices
$n_s$ and $n_t$, both of which are small numbers:
\be 
n_s-1 \equiv k \frac{\d }{\d k} \log \left( \frac{H^4}{\dot \phi^2} \right) = \frac{\d}{H \d t_*} \log \left( \frac{H^4}{\dot \phi^2} \right) = 2(\eta -3\epsilon) \, ,
\ee
and
\be \label{consist}
n_t \equiv k \frac{\d }{\d k} \log \left( H^2 \right) = -2 \epsilon
\ee
In inflationary models, the gravitational waves have a red spectrum, which means $n_t <0 $. A
blue spectrum is only possible if the Null Energy Condition for matter is violated or some
non-standard sound speed is introduced \footnote{The fact that inflation leads to a red tilt
of the tensor power spectrum leads to a way to falsify inflation. There are other cosmological
scenarios such as ``string gas cosmology" \cite{BV,SGCrev} which predicts a blue
spectrum of gravitational waves \cite{BNPV2}}. Cosmic microwave (CMB) anisotropy
experiments can measure $n_s$, and current data indicate that  $n_s - 1$ is slightly negative.

\subsection{Gauges}

In this subsection, we compare different gauges and the resulting different
forms of the scalar perturbations during inflation, and we suggest that the comoving gauge
is the best gauge to compute IR loops. Observables are gauge-invariant and the physics
does not depend on the gauge choice. However, in terms of being able to reliably compute
the observables, there often exists a choice of gauge which is particularly convenient.

Two other gauges besides the comoving gauge have often been used in recent loop
analyses of inflation. The first is a gauge which arises in the effective field theory of
inflation in which one uses a Goldstone modes of gravity perturbations
(see e.g. $\pi$~\cite{Cheung:2007st}). However, for dealing with IR loops this gauge
has problems. The main idea of $\pi$ effective field theory is to apply the Goldstone
equivalence principle to the inflationary perturbation theory. At high energy, the mixing
between Goldstone mode and metric perturbation modes is not important, and then the
scalar perturbation can be treated as a Goldstone boson $\pi$. For single field inflation
models, this effective field theory breaks down for modes whose energy corresponds to
$E < \epsilon^{1/2}H$, $\pi$. But a large contribution to IR loops comes from modes
in this range.

A second frequently used gauge is the uniform density gauge $\zeta=0$. In this case,
the fluctuations are carried by the scalar field perturbation $\delta \phi$ which is non-zero. It is well
known that $\delta \phi$ is not conserved outside horizon. To calculate the tree level two
point function of curvature perturbation, it is convenient to make use of the fact that
$\zeta$ is conserved after crossing the Hubble horizon. Thus, it can be calculated at the
Hubble crossing time. With the relationship matter fluctuations $\delta \phi$ and
metric fluctuations $\zeta$ at this time (valid in any gauge), we can obtain the curvature
perturbations while working in the uniform density gauge. However, in this gauge the IR loop
calculation needs to take into account the time evolution of the integrands on
super-Hubble scales, and thus involves an integration over the whole time of inflation.
In other words, if we use $\delta \phi$,  we must take into account the time dependence
of $\delta \phi$ after Hubble crossing time.

On the other hand, working in comoving gauge in the single field model of inflation, the
curvature perturbation $\zeta$ is conserved outside of the Hubble radius. The
time dependence of correlation functions of the fluctuation variables in this gauge is
trivial.
When we perform the time integration in loop integrals, the IR modes of $\zeta$ are
time-independent and can be taken out of the time integral. Thus, the computations
are technically easier. Note, however, that the small tilt from the k-dependence of
$\zeta$ is essential for the IR loop diagram integration. In the UV, the tilt of the
spectrum is not crucial, since deep inside the Hubble radius the WKB method can be
used~\cite{Weinberg:2010wq}.

To summarize, we conclude that the comoving gauge is the best one for computing
IR loops in single field inflation models. In multi-field models of inflation and in the
presence of non-vanishing entropy perturbations, $\zeta$ is no longer conserved.
The IR divergences of inflation becomes more interesting as we will discuss it in the next
section.

\subsection{Interactions}
\label{interactions}

After choosing the gauge, the next step is to gather the interaction terms for loop
calculation. In comoving gauge, there is no interactive ghost field and thus we need
not consider any ghost terms. In single field inflation with sound speed $c_s=1$,  we
use $\epsilon$ as second expansion parameter for the gravitational action.
The first expansion parameter is the amplitude of the linear fluctuation $\zeta$,
and for each order in $\zeta$ we can expand the corresponding terms in the
interaction Lagrangian in terms of $\epsilon$. Here we just keep
the leading terms in $\epsilon$ for every kind of interaction.
In addition, there are terms proportional to $\frac{\delta \mathcal{L}}{\delta \zeta}$
which can be dealt with by redefinition of $\zeta$ or $h_{ij}$. These terms are neglected
here because they have nothing to do with the leading IR divergences.

In this section, we will list all interaction terms that are relevant to one loop corrections
to two-point function, and we also write down the higher interaction terms since they
can be used to estimate the order of magnitude of the two loop corrections to two-point
functions, or the loop corrections to $n$-point correlation function. In the second
part of the section, we will give the method to estimate the result of the loop corrections.
In the following, we will use conformal time because it is convenient in order to estimate
the order of the loop corrections.

The second order action is $\mathcal{O} (\epsilon)$ and takes the form
\cite{Sasaki, Mukh, Mukhanov:1990me}
\be
S_2 = \int \d \eta  \d x^3 \epsilon a^2 \left( { \zeta ^\prime }^2 -(\partial \zeta)^2 \right)
\ee

The cubic interaction term was first calculated in \cite{Malda}. It is $\mathcal{O} (\epsilon^2)$
and is given by
\be
S_3 = \int \d \eta  \d x^3  \epsilon^2 a^2 \left[   { \zeta^\prime }^2 \zeta - (\partial \zeta)^2\zeta  - 2  \zeta^{\prime} \partial_i \chi \partial_i \zeta  \right] \, ,
\ee
where
\be
  \partial^2 \chi =   \zeta^\prime  \ .
\ee
The interaction Hamiltonian is written as
\be
H_3^{(I)} = -\mathcal{L}_3 \, .
\ee

The leading fourth-order term is of the same order as the cubic interaction $\mathcal{O} (\epsilon^2)$
(it was first computed in \cite{Sloth3})
\be
S_4 = \int \d t \d x^3  \epsilon^2  a^2 \left\lbrace -\beta_j \partial^2 \beta_j - \frac{1}{a H} \partial^{-2} \Sigma  \left[{\zeta^\prime}^2 + (\partial \zeta)^2 \right]   -3 (\partial^{-2} \Sigma)^2 -4  {\zeta^\prime} \beta_j \partial_j\zeta  \right\rbrace \, ,
\ee
where
\be
\frac{1}{2 } \partial^4 \beta_i \equiv \delta^{rs} \left( \partial_i \Sigma_{rs} - \partial_{(r}\Sigma_{s)i} \right)
\ee
and
\be
\Sigma_{rs} \equiv \partial_r {\zeta^\prime} \partial_s \zeta + {\zeta^\prime} \partial_{rs} \zeta \, .
\ee
We have also used $\Sigma \equiv \delta^{rs} \Sigma_{rs}$.
If keep the leading order, the interaction Hamiltonian is also the minus Lagrangian,
\be
H_4^{(I)}= - \mathcal{L}_4
\ee

The fifth order and sixth order terms are more and more complicated, and we just mention the
order in $\epsilon$ of the leading terms in these interaction terms, and we omit partial
derivatives on $\zeta$ in the expressions (this expansion was first discussed
in \cite{Sloth4})
\be
S_5 \sim \int \d t \d x^3 a \epsilon^3 \zeta^5
\ee
and
\be
S_6 \sim \int \d t \d x^3 a \epsilon^3 \zeta^6 \, .
\ee

For the loop corrections, $S_3$ is subdominant compared to $S_4$. The one loop diagram with
$S_3$ interactions has one more vertex than the one with $S_4$. It is thus suppressed by
$\epsilon^2$, but it has one more propagator, which gives an enhancement of $1/\epsilon$.
Therefore, $S_3$ is subdominant.

To compute loop diagrams we must identify the vertex factors and propagators.
Concerning the vertex factors,
\be
S_4 \ \ vertex \mapsto \frac{\epsilon^2}{H^2} \
\ee
and
\be
S_3 \ \  vertex \mapsto \frac{\epsilon^2}{H^2} \ .
\ee
Every internal and external curvature $\zeta$ propagator gives
\be
\zeta \ \  propagator \mapsto \frac{H^2}{\epsilon} \ ,
\ee
and the graviton propagator (which will be relevant since there are
interactions between scalar metric fluctuations and gravitons
at higher than quadratic order) is
\be
h \ \  propagator \mapsto H^2 \ ,
\ee
which comes from the action for free gravitons which is

\be
S_{h2}= \frac{1}{8} \int  \d \eta  \d x^3 a^2\left(    {h}_{ij}^{\prime} {h}_{ij}^{\prime} - \partial_l h_{ij} \partial_l h_{ij} \right) \ .
\ee

Now let us consider the individual interaction terms.
The cubic interaction term between two scalars and a graviton is
\be
S_{\zeta 2 h}= \int  \d \eta  \d x^3 \epsilon a^2 h_{ij} \partial_i \zeta \partial_j \zeta  \ .
\ee

The cubic interaction term between one scalar and two graviton is subdominant in $\epsilon$
since it has the form
\be
S_{\zeta h2}= \int  \d \eta  \d x^3 \frac{\epsilon}{2} H a^3 {h}_{ij}^{\prime} {h}_{ij}^\prime \partial^{-2}{\zeta^\prime} \ .
\ee
The quartic interaction term between two scalars and two gravitons is
\be
S_{\zeta2h2}= \int  \d \eta  \d x^3 \epsilon a^2 \left( -\frac{1}{2} h_{il} h_{lj}\partial_i\zeta \partial_j \zeta \right) \ .
\ee

Counting the order of $\epsilon$ is a easy way to know whether a particular interaction
is dominant or not. Let us consider the leading contributions to the two point function
of $\zeta$:
\begin{equation}
\label{graviton3}
 \includegraphics[width=3cm]{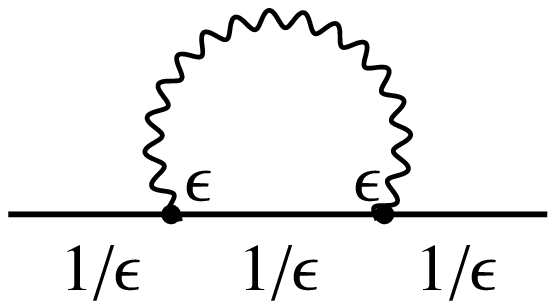} \qquad  \sim \quad \frac{H^4}{\epsilon} \times \left(   \IRdiv  \right)
 \end{equation}

 \begin{equation}
 \label{graviton4}
 \includegraphics[width=3cm]{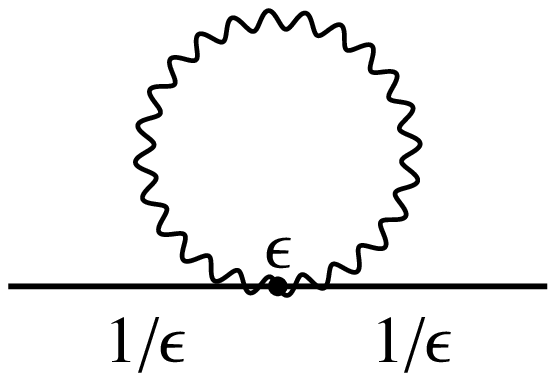}   \qquad  \sim \quad \frac{H^4}{\epsilon} \times \left( \IRdiv  \right)
 \end{equation}

\begin{equation}
 \includegraphics[width=3cm]{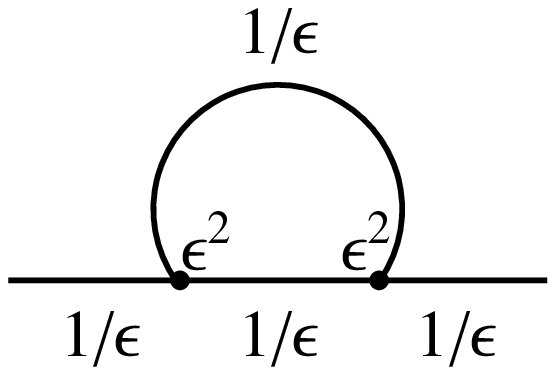} \qquad  \sim \quad H^4\times \left(  \IRdiv  \right)
 \end{equation}

\begin{equation}
\label{curvature4}
 \includegraphics[width=3cm]{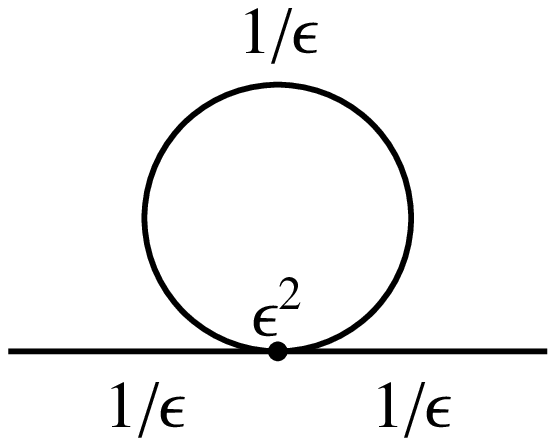}   \qquad  \sim \quad \frac{H^4}{\epsilon} \times \left(  \IRdiv  \right)
 \end{equation}

\begin{equation}
 \includegraphics[width=3cm]{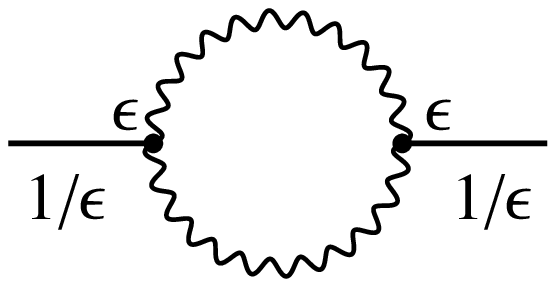}  \qquad  \sim \quad H^4\times \left(  \IRdiv  \right)
 \end{equation}
 \footnote{In fact, because of the derivative terms in the vertex factors, there are no IR divergences in Eq.(\ref{curvature4}).}
 where the wavy lines stand for gravitons and the solid lines are curvature propagators.
 The dominant interaction terms at one loop are $S_4$, $S_{\zeta2h}$ and
$S_{\zeta2h2}$. This counting method can be easily generalized to $n-$point correlation
functions and to higher loop corrections. The IR divergences are logarithmically divergent if
$H$ is constant.

\subsection{In-in Calculation of Loop Corrections}
\label{inin2}

In Sec.~\ref{interactions}, we have obtained the interactions and discussed  the
method to estimate the order of loop diagrams. In this section, we use the
``in-in" formalism to perform the actual calculation of the loop corrections to the
correlation function via evaluating the leading order loop diagrams. As is well
known, in cosmology there is no easily defined ``out" state,
so ``in-in" formalism is applied in order to calculate the observables, e.g. the
correlation functions.

The ``in-in" formalism \cite{Schwinger, Keldysch} is similar to the real-time formalism
in thermal field theory. It makes use of the interaction representation of quantum
field theory. In this formalism, the vacuum expectation value of an operator $\mathcal{O}$
is written as
\be \label{inin}
\left<\Omega| \mathcal{O} (t)   |\Omega  \right>
= \langle 0 \vert\bar{T}e^{i \int_{-\infty}^{t}H_I(t') dt'}
{\cal O}_I (t)   {T}e^{-i \int_{-\infty}^{t}H_I(t') dt'} \vert 0 \rangle
\ee
where $\vert\Omega \rangle$ is the vacuum of the interacting theory and
$\vert 0 \rangle $ is the free field vacuum. The right hand side of the equation
is calculated in the interaction picture.
$T$ and $\bar{T}$ denote time-ordered and anti-time-ordered products.

We are considering the two point function of the curvature perturbation
$\zeta$. There are two kinds of quantum corrections contributing to this
Green's function, one of which is from curvature loops
and the other from graviton loops. The quantum correction from eq~(\ref{graviton3}) is
denoted as $G_p^{I}(\eta) $, and the one from eq.~(\ref{graviton4}) is denoted as
$G_p^{II}(\eta) $. Using the Feynman rules discussed in the previous subsection
they are given by
\bqa\label{digra2}
G_p^{II}(\eta) &=&  2 \epsilon\int_{-\infty}^{\eta}d \eta_1 a^2(\eta_1) {\bf Im} \left[\zeta_p^2(\eta) {\zeta_p^*}^2(\eta_1)
\right]
\int \frac{d^3 q}{(2\pi)^3 } \ 2 p^2 {\rm sin}^2 \theta |h_q(\eta_1)|^2
\eqa
and
\bqa
G_p^{II}(\eta) &=& -\frac{H^2}{4\epsilon p^3} \frac{8}{16 \pi^2} \int_{\Lambda_{IR}} \frac{\d q}{q} H_{q}^2 \ ,
\eqa
where $\theta$ is taken to be the angle between $p$ and $q$.

If $H_{k}$ is not $k$ dependent, the final result will be logarithmically divergent.
However, in inflationary cosmology we must take into account  the small $k$ dependence
and study how this will impact on the physics. Focusing first on $G_p^{II}$ we find
\bqa
\label{Gp2}
G_p^{II}(\eta) &=& \frac{H^2}{4\epsilon p^3} \frac{8}{16 \pi^2} \int_{\Lambda_{IR}} \frac{\d q}{q} H_{initial}^2\left(   \frac{q}{k_{initial}} \right)^{n_t} \nonumber\\
                     &\simeq&  \frac{H^2}{4\epsilon p^3} \frac{1}{16 \pi^2} \frac{8 H_{initial}^2}{ n_t}  \left(   \frac{\Lambda_{IR}}{k_{initial}} \right)^{n_t} \nonumber\\
                     &=&  -\frac{H^2}{4\epsilon p^3} \frac{1}{16 \pi^2} \frac{4 H_{initial}^2}{ \epsilon} \ ,
\eqa
where in the final step we have identified the IR cutoff length scale with the value of the Hubble
radius at the beginning of the inflationary phase. Taking the IR cutoff momentum to be smaller
would make our result probe the state of modes which are super-Hubble throughout the
cosmological evolution.  We have also made use of the relation (\ref{consist})
between $n_t$ and $\epsilon$.

With the same logic and method, the term $G_p^{I}(\eta) $ can be calculated as follows:
\bqa\label{Gp1_c}
G_p^{I}(\eta) &=& -16\epsilon^2 \  {\bf Re} \int_{-\infty}^{\eta}d\eta_1 a^2(\eta_1)
\int_{-\infty}^{\eta}d\eta_2 a^2(\eta_2)
\int d^3 q\  p^4 {\rm sin}^4 \theta \nonumber\\
 &&\times \Big[\theta (\eta_1-\eta_2) \zeta_p^2(\eta) \zeta_p^*(\eta_1) \zeta_p^*(\eta_2)  \zeta_{p'}(\eta_1) \zeta_{p'}^*(\eta_2)
h_q(\eta_1) h_q^*(\eta_2)\nonumber\\
 &&-\frac{1}{2}  |\zeta_p(\eta)|^2  \zeta_p(\eta_1) \zeta_p^*(\eta_2)  \zeta_{p'}(\eta_1) \zeta_{p'}^*(\eta_2) h_q(\eta_1)
h_q^*(\eta_2)\Big] \, ,
\eqa
which leads to
\bqa
G_p^{I}(\eta) &\simeq& \frac{H^2}{4\epsilon p^3} \frac{1}{16 \pi^2} \left( -\frac{4 H_{initial}^2}{ n_t} - \frac{4 H_{initial}^2}{ n_s-1}\right) \nonumber\\
                &=& \frac{H^2}{4\epsilon p^3} \frac{1}{16 \pi^2} \left( \frac{2 H_{initial}^2}{ \epsilon } +\frac{2 H_{initial}^2}{ 3\epsilon -\eta }\right) \, .
\eqa

The one loop corrections from $S_4$ do not have any IR divergences, because the
terms associated with loops are always have a derivative acting on them. Hence,
there are no integrals such as $\int \d q^3 /q^3$ which result, and hence there is
no IR divergence.

Notice that $G_p^{I}$ and $G_p^{II}$ have different signs, and thus there is a
possibility that the terms might cancel. In fact, if we consider a theory in
which $\epsilon$ is very slowly varying on a time scale of $H^{-1}$ compared
to its actual value, which often happens in models which
admit eternal inflation or in which inflation simply lasts a lot longer than the
minimal 60 e-foldings which it needs to last in order to be successful at
resolving the flatness and horizon problems of Standard Cosmology, then
\be
\eta \simeq 2\epsilon \
\ee
and the leading order IR divergences is cancelled.
On the other hand, if $\frac{\dot \epsilon} {H \epsilon}$ is of the same order as
$\epsilon$, then the analysis is more complicated, and we need to take into account
the detailed time dependence of $\epsilon$.

In models in which inflation only lasts
60 e-folding the leading IR divergences do not cancel. However,
the magnitude of the loop corrections is small because the phase space of
IR modes is small. 

\section{Entropy Modes}
\label{dotzetanonzero}

The restriction to single field models of inflation is not very physical. We know that
matter in the real world is described by a large number of fields. The inflaton is
one additional field which must be added to the theory. The inflaton must also
couple to other fields in order to provide a successful exit from the inflationary
phase. The presence of other matter fields leads to the presence of entropy
modes. Thus, when computing quantum effects on cosmological observables
it is essential to take into account the effects of loops involving entropy modes.

Whereas is some circumstances adiabatic loops lead to effects which can be
gauged away (as we have seen in the previous section),  effects of entropy
loops cannot (see e.g. \cite{Ghazal2}). Thus, if the cosmological fluctuations
can be influenced by IR effects from entropy loops, these effects cannot be gauged away.
They will contribute to the effective field theory of inflationary perturbations and
can hence influence cosmological observations. In this section, we will study the
IR divergences to curvature perturbation via entropy loops. We will see that these
effects can be very important and that these loops effects can lead to a breakdown
of perturbation theory if the self-coupling constant of the entropy modes is larger
than some critical value which is much smaller than $1$. The magnitude of
the infrared effects of entropy modes depends on the mass of the entropy modes
and on the initial conditions. We separate this section into three parts, one of
which deals with massless entropy modes, and the second subsection considers
entropy modes with a small mass. These two cases will give different IR divergences.
And in the final part we will deal with a more general case in which we take the 
Higgs field to be the inflaton as a example.

Several groups have studied entropy loops~\cite{Weinberg:2005vy,Seery:2010kh} by
adding a (approximately) massless scalar field $\sigma$ to the matter sector which
minimally couples to gravity. The Lagrangian for $\sigma$ is
\be
\mathcal{L} = -\frac{1}{2} \partial_\mu \sigma \partial^\mu \sigma \, .
\ee
The correlation function $\langle\zeta \zeta \rangle$ does not pick up important
IR contributions from the kinetic term, since the derivatives will introduce an extra
power of momentum into the loop integrals which make the integrals more
UV rather than IR divergent. Considering the simplest interaction between the
inflaton and the entropy field,
\be
\mathcal{L} = -\frac{1}{2} \partial_\mu \sigma \partial^\mu \sigma  - g \phi^2 \sigma^2 \ .
\ee
it appears that there are sizeable IR divergences. However, unless $g$ is very small, the
entropy field $\sigma$ will acquire a large mass from the expectation value of the
background inflaton field. The effective IR cutoff for entropy loops is thus the larger of either
the Hubble scale (the $\sigma$ mass induced by gravitational effects) or the induced mass
of the entropy mode. Thus, the IR contribution of entropy modes is tiny.

However, the conclusion that the entropy IR modes contribute little to the two point
correlation function of $\zeta$ has an important loophole, which relies on the
fact that the curvature perturbation is no longer constant outside Hubble radius.
It is well-known that entropy fluctuations source a growing curvature perturbation.
Also, the entropy self-interaction term will contribute to the curvature perturbations
via the gravitational coupling. In order to study the effects of entropy perturbations,
there are two ways to proceed. The first is to solve for the time dependent curvature
perturbations from the equation of motion for $\zeta$ in the presence of entropy
sources and then to perform the loop calculation using the in-in formalism.
The second approach (the one we now adopt), is to solve the free equation of motion
for $\zeta$, thus obtaining a time-independent solution, and then treat the source
term as an interaction term. We then use the in-in formalism to obtain the
corrections from IR entropy modes \footnote{The corresponding result is equivalent
to the leading order in expansion with respect to the coupling between entropic and
adiabatic modes of the result in the first approach.  This perturbative approach
is employed in \cite{0908.4035} in evaluating the cross-correlation between
adiabatic and entropic mode and in \cite{0909.0496} and \cite{YiXingang} in evaluating the
transfer of non-Gaussianities from the entropic to the adiabatic sector.}.

 \subsection{Massless Entropy Modes}

It is a general result that there is mixing between isocurvature and curvature modes.
One simple example in which it is easy to see this mixing is to consider a model
with a complex scalar field with a ``Mexican hat" potential. The adiabatic mode
corresponds to the scalar field moving in angular direction in the vacuum
manifold, and the isocurvature direction is along the radius.
The action describing the system is as follows \cite{YiXingang}
\be{\label{S_2fields}}
S =  \int \d \eta  \d ^3 x \frac{1}{2} \left(  {a^2} {\theta^\prime}^2 (R+\sigma)^2 - a^4\lambda \sigma^4 \right) \ ,
\ee
where $R$ represents the radius of the circle in field space, and $\theta$ is the
angle of the scalar field configuration. The field $\theta$ represents the curvature
perturbation. The self interaction of the entropy modes can have various form; here
we use a $\lambda \sigma^4$ interaction without loss of generality. If the
interaction does not contain more than two time-like or spacial derivatives, then
entropy loops lead to a IR divergence.

The mixing term is treated as an interaction. As we argued in Section 2, it is
convenient to work in the gauge given by $\delta \theta =0$. To obtain the
interaction Hamiltonian in this gauge, we start with the perturbative expansion
of (\ref{S_2fields}) and then (valid to first order approximation) transform
the interaction term to the comoving gauge by the transformation
\be
R \delta \theta \rightarrow \zeta (- \dot\phi /H) \, .
\ee

The interaction Hamiltonian then becomes\footnote{ In multi-field models, in
general there are two types of couplings between adiabatic and entropic modes
(see e.g. \cite{0801.1085}): $\sim \xi \zeta'\sigma$ and $\sim \mathcal{H}\xi\zeta\sigma$
with the common coefficient $\xi$ denoting the tuning rate in field space
(as $\dot{\theta}$ in eq.(\ref{S_2fields})). In this work, in order to see the
qualitative effect of entropic mode, we only take into account the contribution
 from the first type.
}
\be
\label{2pinteraction}
\mathcal{H}_I = -2 a^2  {\theta}_0^\prime  \delta{\theta}^\prime R \sigma  \rightarrow  4 a^3 \frac{\epsilon H}{R}  \zeta^{\prime} \sigma
\ee
where the second arrow indicates the result after performing the gauge
transformation. If the entropy modes are massless, we have the exact
solution of the free entropy modes which start on sub-Hubble scales
in their vacuum state:
\be
\sigma =  \frac{H_*}{ \sqrt{2 k^3} } ( 1+ i k \eta) e^{-i k\eta} \, .
\ee
Note that these mode functions freeze out and become constant in time
on super-Hubble scales.

First, the time dependent part of tree level contribution to the two
point function \footnote{See \cite{0908.4035} for a related calculation
of the corrections to the adiabatic power spectrum due to the coupling
between adiabatic and entropic modes.} is captured by the two point mixing
interaction term~(\ref{2pinteraction}) and yields
\bqa
\label{treelevelcorrection}
 G_{tree} &=&  - 2 \Re \langle \int H_I \int H_I \zeta \zeta \rangle +  \langle \int H_I \zeta \zeta \int H_I \rangle \nonumber\\
      &=& (2\pi)^3 \delta ^3  \left( \textbf{k}-\textbf{k}^\prime \right)  \left(  \frac{4 \epsilon H} {R} \right)^2
\nonumber\\ &&
      \left[ -4\Re \int^{\eta}_{-\infty} \d \eta_2 a^3(\eta_2)  \int^{\eta_2}_{-\infty} \d \eta_1 a^3(\eta_1)   \sigma_k(\eta_1) \zeta^{\prime}_k (\eta_1) \sigma_k^*(\eta_2) \zeta^{\prime}_k (\eta_2) \zeta^*_k (\eta) \zeta^*_k(\eta)  \right.\nonumber
      \\
       && \left.
      +2 \int^{\eta}_{-\infty} \d \eta_1 a^3(\eta_1)  \int^{\eta}_{-\infty} \d \eta_2 a^3(\eta_2) \sigma_k(\eta_1) \zeta^{\prime}_k (\eta_1) \zeta^*_k (\eta) \zeta_k(\eta)  \sigma_k^*(\eta_2) {\zeta^{\prime}_k}^*(\eta_2) \right] \nonumber\\
      &=& (2\pi)^3 \delta ^3  \left( \textbf{k}-\textbf{k}^\prime \right) \frac{8 \epsilon M_{pl}^2}{R^2} \frac{H^2}{4\epsilon k^3} \times \left[ \log \left(-k\eta \right) \right]^2
\eqa

The tree level correction from the entropy modes is given by the following diagram
and yields
\begin{equation}
\label{treeapp}
 \includegraphics[width=5cm]{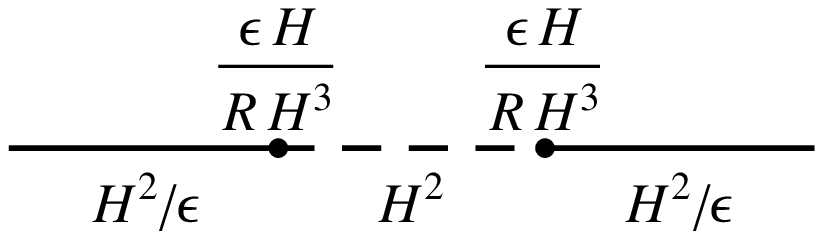}  \qquad \sim \quad \frac{\epsilon}{R^2} \frac{H^2}{\epsilon} \times \left[ \log \left(-k\eta \right) \right]^2 \, .
 \end{equation}
Here, the solid lines denote curvature correlations
$\langle \zeta \zeta \rangle \sim \frac{H^2}{\epsilon}$, and the
dotted lines are entropy modes with propagator
$ \langle \sigma \sigma \rangle \sim H^2$. The interaction vertex
is proportional to $\frac{\epsilon H}{R}$.
In this estimate, the extra $H^3$ in the denominators of the interaction terms
arise from the H-dependence of the scale factor $a^3=-\frac{1}{H^3\eta^3}$
arising in the interactions.
The entropy perturbations source the curvature perturbation via
\be
{\dot{\zeta}} \sim  \delta s \, ,
\ee
and this explains why the tree level result above contains
the two factors of $\log \left(-k\eta \right) \sim t$.

So far, we have neglected the mass and self-interactions of the entropy field.
For the moment, we will assume that the mass of the entropy modes is small and
compute the effects of the self-interactions. To be specific, we choose a
$\sigma ^4 $ interaction
\be
\mathcal{H}_I= a^4\lambda \sigma^4
\ee
and will go on to study the resulting entropy loop correction to the
curvature two point function. It is given by the following Feynman
diagram
\bqa
\label{loopestimate}
 \includegraphics[width=4cm]{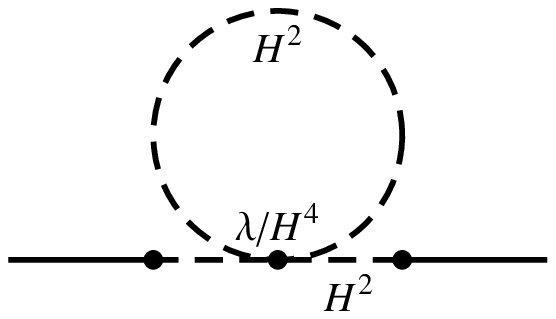}
	       &\sim &
 \includegraphics[width=4cm]{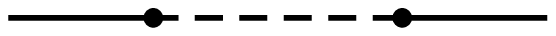} 	
	      \times\lambda \log (-k\eta) \times (\IRdiv) \nonumber \\
	         &\sim &   \frac{\epsilon}{R^2} \frac{H^2}{\epsilon} \times \left[ \log \left(-k\eta \right) \right]^2 \times \frac{\lambda}{H^2} \frac{H^2_{initial}}{\epsilon} \times \log \left(-k\eta \right) \, .
\eqa
As in Section \ref{inin2} (Eq. (\ref{Gp2})), the power-law infrared divergence 
leads to the factor of $H^2_{initial} / \epsilon$, where the
IR cutoff length scale is set to be the Hubble radius at the beginning of inflation.
The physical reason for having the extra term $\log (-k\eta) $ is that the
entropy mode also sources itself and the effect increases with time.
The IR divergence is expresed via the term $ \frac{H^2_{initial}}{\epsilon}$
where the IR cutoff length scale is (as in Section 2) set to be the Hubble radius
at the beginning of inflation. This divergence is similar in structure to what
we found in Eq.~\ref{Gp2}. However, in this case there can be no cancellation in
the limit of very slow-roll inflation. Note that the logarithm divergences of pure
de Sitter space has changed to a power law divergence. Summarizing, the entropy loop
correction to the curvature two point function is given by
\bqa
\label{massiveloop}
 G_{loop} &=&  2 \Im \langle \int H_I \int H_I\int H_I \zeta \zeta \rangle -2 \Im \langle \int H_I \int H_I \zeta \zeta \int H_I \rangle \nonumber\\
&=& (2\pi)^3 \delta ^3  \left( \textbf{k}-\textbf{k}^\prime \right)  \lambda \frac{14 M_{pl}^2}{3 \pi^2 R^2} \frac{H^2}{4\epsilon k^3} \left( \frac{H_{initial}}{H} \right)^2  \times \left[ \log \left(-k\eta \right) \right]^3 \, .
\eqa

Comparing the tree level result with the one loop contribution we see that
perturbation theory breaks down when
\be
\lambda  \frac{7}{12\pi^2} \frac{1}{\epsilon}
\left( \frac{H_{initial}}{H} \right)^{2} \times \left[ \log \left(-k\eta \right) \right]
\, > \, 1 \, .
\ee
Using dynamical renormalization group techniques \cite{Burgess:2009bs}, the
temporal logarithm divergences $\left[ \log \left(-k\eta \right) \right] $ can be
resummed. In fact, one way to interpret this result is that interaction generate a
mass of order $\frac{\lambda}{\epsilon}  \left( \frac{H_{initial}}{H} \right)^2$.

After doing this, the condition for the breakdown of perturbation theory
changed slightly and becomes
\be
\lambda  \frac{1}{4\pi^2} \frac{1}{\epsilon}  \left( \frac{H_{initial}}{H} \right)^{2}
\, > \, 1 \, .
\ee
which comes from the subleading term in the loop corrections
and is still sensitive to the initial condition.
This constraint on $\lambda$ is severe, in particular in models with
a very long period of inflation when the ratio of $H$ values is large.

\subsection{Massive Entropy Modes}

Massive entropy modes are not constant outside the Hubble radius (even at
level of free field theory). The free field equation of motion is
\be
\sigma_k^{\prime \prime} -\frac{2}{\eta}\sigma_k^{\prime } + \left(  k^2 + \frac{m^2}{H^2 \eta^2} \right) \sigma_k=0 \, .
\ee
After choosing the Bunch-Davis vacuum at past infinity the solutions are
\be
\sigma_k = - i \e {i(\nu + \frac{1}{2})\frac{\pi}{2} } \frac{ \sqrt{\pi}}{2} H \left( -\eta \right)^{\frac{3}{2}} \mathrm{H}_\nu^{(1)}\left( -k\eta \right)
\quad \quad  {\rm for} \quad \frac{m^2}{H^2} <\frac{4}{9} \ ,
\ee
where $H_{\nu}$ indicates a Hankel function and
\be
\nu= \sqrt{\frac{9}{4} -\frac{m^2}{H^2} } \, .
\ee
If $ -k\eta  \ll 1 $ (super-Hubble scales), the values of the modes are
slowly decreasing in time:
\be
\sigma_k  \simeq - \e {i(\nu + \frac{1}{2})\frac{\pi}{2} } \frac{2^{\nu -1}} {\sqrt{\pi}} \Gamma(\nu) \frac{H}{k^{\frac{3}{2}}} \left( -k\eta \right)^{-\nu +\frac{3}{2}}
\ee
In the case of $ m \ll H $ we have $ \nu \simeq \frac{3}{2} -\frac{1}{3}\frac{m^2}{H^2} $,
and hence
\be
\left(-\eta \right)^{-\nu +\frac{3}{2}} \simeq \left( a(\eta) H \right)^{-\frac{1}{3}\frac{m^2}{H^2}   } \simeq \e { -\frac{1}{3}\frac{m^2}{H^2} \left(H t
+ \ln H \right) }
\ee
Thus the entropy modes will have decayed away on a time scale of
$ H t \sim 3 \frac{H^2}{m^2} $. Assuming that $m \sim 10^{-2} H$,
then the decay will have occurred after $10^4$ e-foldings. Thus, the
effective IR cutoff for the massive entropy modes is
\be
\Lambda_{IR} \sim  H \e {-3 \frac{H^2}{m^2}} \, .
\ee
If inflation does not last as long as $ 3 \frac{H^2}{m^2}$ e-foldings, the
IR cutoff is the same as for massless entropy modes when it is set by
the initial value of the Hubble radius, as we have previously discussed.

We now first calculate the tree level correction to the curvature
correlation function. The calculation is similar to the analysis of
Eq.~(\ref{treelevelcorrection}) in the case of  massless entropy.
We will just write down the leading order terms:
\bqa
G_{tree} &=& (2\pi)^3 \delta ^3  \left( \textbf{k}-\textbf{k}^\prime \right)  \left( \frac{4 \epsilon H} {R} \right)^2   \times \left[
\frac{\left(   -k\eta \right)^{-\nu+\frac{3}{2}} }{ 4\epsilon^2 \left( \nu -\frac{3}{2} \right)^2 k^3 }  \right. \nonumber\\
&& \left.- \frac{ 4^{\nu-1} \left(    \cos (2\pi \nu )   -\cos (4\pi \nu) \right) \Gamma (\nu)^2 }{ 2\pi^2 k^3 \epsilon^2 \left(2\nu-1 \right)} \left(   -k\eta \right)^{-2\nu+1}\right]
\eqa
The first term in the bracket is the same as Eq.~(\ref{treelevelcorrection}),
with the $\log ^2 (-k\eta)$ divergence being recoved in the limit of
$\nu\rightarrow \frac{3}{2}$, and the second term is zero in this limit.

The one loop result is complicated, and we here only show the order of magnitude of the
result which can be derived from the in-in formalism similar to in Eq.~(\ref{massiveloop}),
or can be estimated as in Eq.~(\ref{loopestimate}). The result is
\bqa
G_{loop} \sim G_{tree} \times \lambda   \frac{1}{\epsilon} \left(\frac{H_{initial}}{ H} \right)^2 \frac{\left(    -k \eta \right)^{x\left(-\frac{3}{2}+\nu \right)}}{\nu-\frac{3}{2}}
\eqa
where $x$ is a non-negative integer of the order one. Comparing with the result
for massless entropy modes, we see that the logarithm divergence in time becomes
a power law divergence.

Comparing the magnitudes of the tree level and one loop terms again leads to a
condition on $\lambda$ in order that perturbation theory remain valid.
The most conservative constraint on the coupling constant $\lambda$, independent
whether the entropy mode is massless or massive, is
\be \label{limit1}
\lambda < \epsilon \e {-2 N \epsilon} \frac{1}{N^{\prime}}
\ee
where $N$ is the total e-folding number of the inflationary period, and
$N^{\prime}$ denotes how many e-foldings before the end of inflation that the
$k$ mode corresponding to the external line of the curvature perturbation exits
the Hubble radius. Choosing the fiducial values $N\sim 60$, $\epsilon \sim 0.1$
and $N^{\prime} \sim \mathcal{O} (1 \div10)$, then
\be \label{limit2}
\lambda < 10^{-7} \div 10^{-6}
\ee
However, in large field models of inflation the value of $N$ is much larger than
the minimal value chosen to obtain the above estimate, and then the constraint
(\ref{limit1}) will be much more stringent than the result (\ref{limit2}).

\subsection{Two Loops and Higgs Field}

In this subsection, we will discuss a more general situation which is expected to
occur during inflation. Besides the inflaton, we expect physics to contain many
other scalar fields, such as the Higgs field(s), moduli fields, and superpartners of
fermions. There is the possibility that these scalars do not directly interact
with the inflaton but only interact with curvature perturbations via gravity.
If the masses of these scalar fields are light compared to the Hubble scale (which we
expect to be true in the case of the Higgs field(s)), then the modes of these
scalar field will decay at $3\frac{H^2}{m^2}$ e-foldings after Hubble radius crossing,
as discussed above. For example, the decay of the Higgs field with mass
$\mathcal{O}(100) GeV$ begins at $\frac{H^2}{m^2} \sim  \mathcal{O} (10^{25})$
e-foldings after the modes leave the Hubble horizon. Therefore, for fields with
a small mass, the IR cutoff will usually be given by the initial Hubble radius
at the beginning of the inflationary phase.

The interactions terms for the scalars come from self-interactions
\bqa
S_I &=&  \int \d \eta  \d ^3 x  N \sqrt{h}  \left (- \lambda \sigma^4 \right) =  \int \d \eta  \d ^3 x  a^4 \e {3 \zeta + \zeta^\prime /(a H )} \left (- \lambda \sigma^4 \right) \nonumber\\
    & \supset &- \int \d \eta  \d ^3 x   a^4 \left(  \frac{9}{2} \zeta^2 + \frac{3}{a H} \zeta \zeta^\prime + \frac{1}{2 a^2 H^2} {\zeta^\prime}^2  \right) \lambda  \sigma^4
\eqa
Here we neglect the interactions with the graviton, and only consider the
interaction with the curvature perturbation $\zeta$, since the two point
correction function of curvature perturbations are considered.
This kind of interaction can lead to two types of loop correction to the
two point function of $\zeta$.

We can estimate the magnitude of the loop corrections
by counting powers in $H$ and $\epsilon$ and making use of the
power-law IR divergences which depend on the initial value of the
Hubble scale, as in previous subsection. The Feynman diagram
and the scaling of its contribution to the two point function is
\bqa
 \includegraphics[width=4cm]{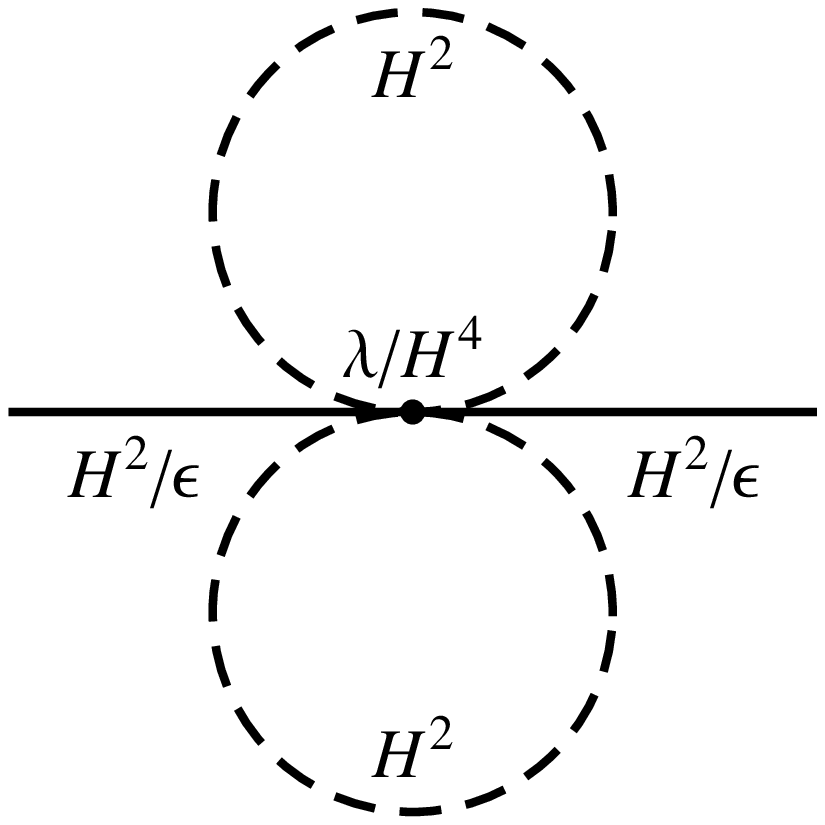}   &\sim & \lambda\frac{H^4}{\epsilon^2} \times \left(  \IRdiv  \right) \nonumber\\
	                             &\sim &  \lambda \frac{H^2}{\epsilon} \times \frac{H_{initial}^4}{\epsilon^3 H^2 M_{pl}^2} \, .
 \eqa
The magnitude of the loop correction can be calculated by in-in formalism and yields
\bqa
\label{2loop}
 G_{2 loop} &=& - 2 \Im \langle \int H_I  \zeta \zeta \rangle  \nonumber\\
&=& (2\pi)^3 \delta ^3  \left( \textbf{k}-\textbf{k}^\prime \right)  \frac{H^2}{2\epsilon k^3}  \frac{3}{8 (2\pi)^4 \epsilon^3} \frac{H_{initial}^2}{H^2} \frac{H_{initial}^2}{M_{p}^2} \times \lambda \, .
\eqa

We can now ask what constraint on the coupling constant results. The answer
depends on how long inflation lasts and how large $H$ is.
If we assume that there was eternal inflation in the early universe, and
that thus the density fluctuations initially were very large (on scales
which are obviously completely irrelevant to current observations), i.e.
 \be
 \left(\frac{\delta \rho}{\rho}  \right)^2 = \frac{H_{initial}^2}{2\epsilon (2\pi)^2 M_p^2}
\geq 1 \, ,
 \ee
then we can derive the following constraint on the coupling constant $\lambda$: If
we require that the two loop correction is not larger than the tree level value, then
\be
\lambda < 10^{-10} \times \epsilon^2 < 10^{-12} \ ,
\ee
where $10^{-10}$ is from the thermal fluctuations of CMB.
Thus, if we want to trust the effective field theory of eternal inflation,
all scalar fields with the small mass must have a self-interaction
coupling coonstant which is smaller than $10^{-12}$.

\section{Discussion}

In this paper, we have computed one loop corrections to the two-point
correlation function of the scalar field $\zeta$ which characterizes the
curvature perturbations. We emphasized that the k-dependence of the Hubble
scale during inflation will change the IR logarithmic divergences of pure de
Sitter space into power law divergences. The power divergence implies that
the IR divergences are sensitive to the initial value of the Hubble
parameter, the natural IR cutoff scale in the context of inflationary
cosmology.

We argued that the best choice of the gauge to perform the computation of
IR looops  in single field inflation models is the comoving gauge, since
in it the perturbation variable (namely $\zeta$) is not time-dependent on
super-Hubble scales.

We showed how to estimate the IR divergences in one loop corrections to the
two point correlation function of $\zeta$, and we also used the in-in formalism
to calculate the one loop diagram, both of which are consistent. The correction
is of the order $\mathcal{O} (\frac{H_{initial}^2}{\epsilon}) $. In
single field models of inflation, then if the inflationary phase
lasts long, the requirement that the slow roll condition on
$\epsilon$ is preserved in time will lead to some cancellations of the loop
corrections, and the semi-classical calculation breaks down only when
$H_{initial}^2 \sim 1$. The cancellation of the leading IR divergences in single
field inflation models is consistent with the semi-classical relation
obtained in \cite{Giddings:2010nc}.

However, in multiple field inflation models, the curvature perturbation
$\zeta$ is not conserved on super-Hubble scales. The entropy modes source
the curvature perturbation and also self-interact. Neither the kinetic
term of the entropy field nor the interaction terms coupling the entropy field
to the inflaton lead to an IR divergence, but the self-interaction terms do lead
to IR divergences. Massless entropy modes will lead to logarithm divergences in
conformal time, while the massive (but low mass) ones will lead to power law
divergences in time. Estimating the ratio of the one loop to tree level results
leads to constraints on the self-coupling of the entropy modes which have
to be satisfied in order to avoid the breakdown of perturbation theory.
In the two fields inflation models, for a 
$\lambda \sigma^4$ interaction term (where $\sigma$ is the
entropy field), the constraint takes the form
$\lambda <  10^{-7} \div 10^{-6}$ if we require that the effective field theory
for inflation remain valid for more than $60$ e-folding. Because this
constraint is on the self-coupling, we believe that it is rather generic.
If inflation lasts longer than the minimal duration it must last in order
to address the cosmological problems of the Standard Big Bang model,
then the constraints on $\lambda$ become much more restrictive.
In more general cases, any scalar (for example, the Higgs field) which 
does not couple directly to the inflaton field, couples to gravity. The leading order correction 
which such a field has on the curvature perturbation is from a two-loop diagram, 
and the requirement that the two-loop correction is not larger than the tree level 
result leads to a constraint of the self-coupling of the scalar field of the form
$\lambda < 10^{-12}$ if the inflaton field starts in the region of eternal inflation. 

Our analysis has implications for the recent debate about the stability of
de Sitter space. There is one camp \cite{Marolf} arguing for the stability of de Sitter
space even in the presence of fluctuations based on performing the
analysis in Euclidean de Sitter space. A second camp \cite{Mottola}
argues that in a cosmological context a de Sitter phase is unstable because
of large IR back-reaction effects. The renormalization prescription breaks
the de Sitter invariance as can be seen either by imposing an initial
space-like Cauchy surface (e.g. the initial time of the inflationary phase),
or by setting the IR cutoff to be fixed in comoving coordinates (see e.g. \cite{Lowe})
for a recent comparison of these two camps). We have shown that in
the presence of entropy fluctuations, the effects which lead the second
camp to argue that de Sitter space is unstable will be much larger in
magnitude. Thus, our analysis may imply that entropy modes lead to
an enhanced decay rate of de Sitter space.

\section*{Acknowledgements}

We would like to thank  L. Leblond, E. Lim, S. Shandera and R. Woodard for 
interesting conversations. X. G. is supported by ANR (Agence Nationale de la 
Recherche) grant ``STR-COSMO" ANR-09-BLAN-0157-01.
W. X. is supported in part by a Schulich Fellowship    ,
and R.B. is supported in part by an NSERC Discovery grant and by funds from
the Canada Research Chair program. R.B. wishes to thank Patrick Peter and Jerome Martin
for an invitation to the IAP during which this
project was finalized. This visit was supported by
the grant ``PICS Cosmologie de l'Univers Primordial"
to the IAP. We wish to thank M. Sloth, A. Riotto, R. Woodard, L. Leblond and 
M. Zaldarriaga for comments on the manuscript.



\end{document}